\documentclass[preprint,3p,11pt]{elsarticle}
\usepackage[T1]{fontenc}

\usepackage{amsmath,amssymb,mathtools,mathrsfs,xspace,subfigure}
\usepackage{amsfonts}
\usepackage[usenames,dvipsnames]{xcolor}
\usepackage[]{graphicx}
\usepackage{url}
\usepackage{hyperref}
\usepackage{fullpage}
\usepackage{url}
\usepackage{float}
\usepackage{bm}
\usepackage{soul}
\usepackage{natbib}
\usepackage{enumitem}

\graphicspath{{./Figures/}}

\begin{document}

\begin{frontmatter}



\title{Discrete Differential Geometry for Simulating Nonlinear Behaviors of Flexible Systems: A Survey}


\author[a]{Dezhong Tong\fnref{label2}}
\author[b]{Andrew Choi\fnref{label2}}
\author[a]{Jiaqi Wang\fnref{label2}}
\author[c]{Weicheng Huang\fnref{label2}\corref{cor5}}
\author[a]{Zexiong Chen}
\author[d]{Jiahao Li}
\author[a]{Xiaonan Huang}
\author[e]{Mingchao Liu\corref{cor5}}
\author[f,g]{Huajian Gao\corref{cor5}}
\author[h,i]{K. Jimmy Hsia\corref{cor5}}

\fntext[label2]{D.T., A.C., J.W., and W.H. contributed equally to this work.}
\cortext[cor5]{Corresponding authors: weicheng.huang@newcastle.ac.uk (W.H.), m.liu.2@bham.ac.uk (M.L.), gao.huajian@tsinghua.edu.cn (H.G.), kjhsia@ntu.edu.sg (K.J.H.)}

\address[a]{The Robotics Department, University of Michigan, Ann Arbor, Michigan, 48109, USA}
\address[b]{Horizon Robotics, Cupertino, CA, 95014, USA}
\address[c]{School of Engineering, Newcastle University, Newcastle upon Tyne NE1 7RU, UK}
\address[d]{CAS Key Laboratory of Mechanical Behavior and Design of Materials, Department of Modern Mechanics, University of Science and Technology of China, Hefei 230027, People's Republic of China}
\address[e]{Department of Mechanical Engineering, University of Birmingham, Birmingham B15 2TT, UK}
\address[f]{AML, Department of Engineering Mechanics, Tsinghua University, Beijing 100084, People's Republic of China}
\address[g]{Mechano-X Institute, Tsinghua University, Beijing 100084, People's Republic of China}
\address[h]{School of Mechanical and Aerospace Engineering, Nanyang Technological University, Singapore 639798, Republic of Singapore}
\address[i]{School of Chemistry, Chemical Engineering and Biotechnology, Nanyang Technological University, Singapore 639798, Republic of Singapore}

\begin{abstract}
Flexible slender structures such as rods, ribbons, plates, and shells exhibit extreme nonlinear responses—bending, twisting, buckling, wrinkling, and self-contact—that defy conventional simulation frameworks. Discrete Differential Geometry (DDG) has emerged as a geometry-first, structure-preserving paradigm for modeling such behaviors. Unlike finite element or mass–spring methods, DDG discretizes geometry rather than governing equations, allowing curvature, twist, and strain to be defined directly on meshes. This approach yields robust large-deformation dynamics, accurate handling of contact, and differentiability essential for inverse design and learning-based control. This review consolidates the rapidly expanding landscape of DDG models across 1D and 2D systems, including discrete elastic rods, ribbons, plates, and shells, as well as multiphysics extensions to contact, magnetic actuation, and fluid–structure interaction. We synthesize applications spanning mechanics of nonlinear instabilities, biological morphogenesis, functional structures and devices, and robotics from manipulation to soft machines. Compared with established approaches, DDG offers a unique balance of geometric fidelity, computational efficiency, and algorithmic differentiability, bridging continuum rigor with real-time, contact-rich performance. We conclude by outlining opportunities for multiphysics coupling, hybrid physics–data pipelines, and scalable GPU-accelerated solvers, and by emphasizing DDG’s role in enabling digital twins, sim-to-real transfer, and intelligent design of next-generation flexible systems.


\end{abstract}

\begin{keyword}
Discrete Differential Geometry \sep Nonlinear Flexible Structures \sep Slender Rods and Shells \sep Simulation and Control \sep Soft Robotics \sep Multiphysics Coupling
\end{keyword}

\end{frontmatter}

\newpage  
\tableofcontents
\newpage

\section{Introduction}
Flexible slender structures—rods, ribbons, plates, and shells—pervade natural and engineered systems, from DNA and plant tendrils to surgical instruments, cables, and soft robots.
Their functionality hinges on nonlinear behaviors such as buckling, wrinkling, and self-contact, which remain difficult to capture with conventional tools~\cite{huang2025tutorial}.
Finite element methods offer accuracy but struggle with extreme deformations and contact~\cite{cremonesi2020state}, while mass–spring and position-based dynamics trade realism for efficiency~\cite{yin2021modeling}. Discrete Differential Geometry (DDG) offers a promising alternative. By discretizing geometry directly, DDG preserves curvature and twist at the mesh level, enabling robust large-deformation simulations with structure-preserving fidelity~\cite{grinspun2006discrete}. Initially rooted in computer graphics, DDG is now migrating into mechanics and robotics, where differentiability and efficiency are crucial for inverse design and control. This review surveys the foundations, representative models, and applications of DDG in nonlinear flexible systems, and outlines future opportunities for integrating DDG into multiphysics, machine learning, and engineering practice.

Across science and technology, slender structures play important roles at vastly different length and time scales~\cite{reis2018mechanics}. 
In biology, they encode and transmit the essential functions of life, from the folding of DNA~\cite{thompson2002supercoiling, manning1996continuum, djurivckovic2013twist, stevens2001simple} and the dynamics of proteins~\cite{benham1977elastic, yang2009protein, marantan2018mechanics} to the morphogenesis of plant tendrils~\cite{moulton2013morphoelastic, lessinnes2017morphoelastic, moulton2020morphoelastic, goriely1998spontaneous}. 
In engineering, they are central to the operation of devices and infrastructures, from tension-bearing components in bridges~\cite{bilisik2015applications,vu2015finite} and aerospace composites~\cite{mahadevan2024knitting, baek2021smooth, jauffres2010discrete, benvenuto2015dynamics} to wearable devices and medical catheters that operate safely within the human body~\cite{li2024model, tong2025real, kim2019ferromagnetic}. 
Their ubiquity arises from a shared set of mechanical advantages: high compliance, adaptability to complex environments, and efficient load transfer. These properties make them essential for modern technologies that demand lightweight, deformable, and responsive designs.

\begin{figure}[h]
    \centering
	\includegraphics[width=\textwidth]{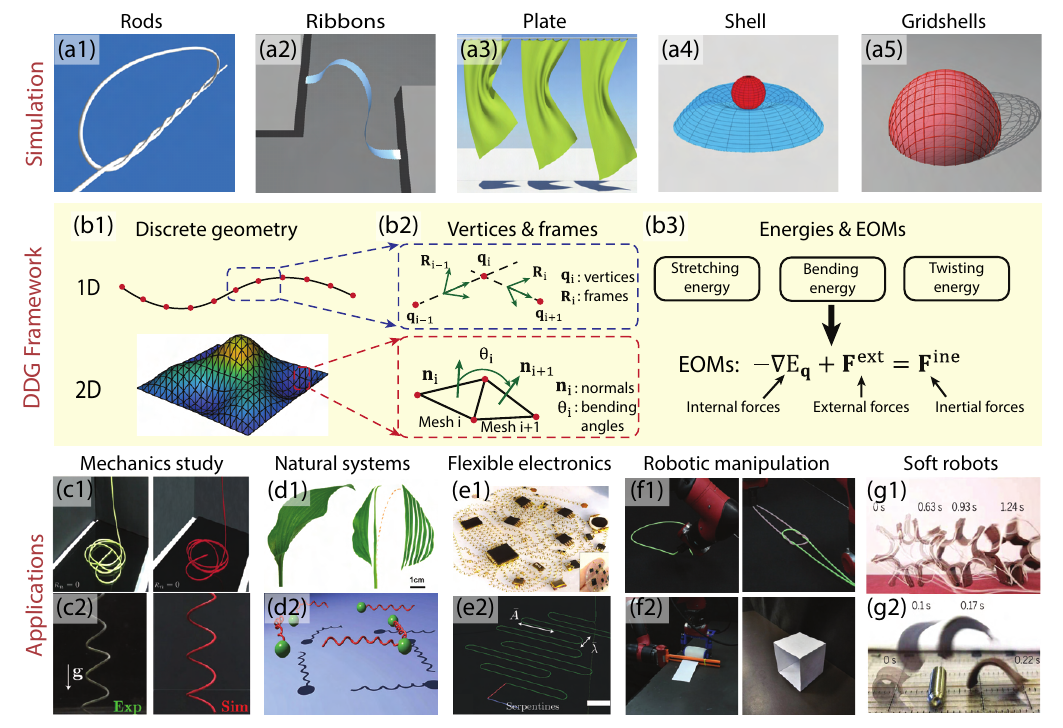}
	\caption{Overview of DDG-based simulations and applications.
    (a) Representative structures: (a1) rods~\cite{choi2021implicit}; (a2) ribbons~\cite{huang2020shear}; (a3) plates~\cite{bouaziz2023projective}; (a4) shells~\cite{huang2024discrete}; (a5) gridshells~\cite{huang2021numerical}.
    (b) DDG framework: (b1) discrete geometry for slender structures in 1D and 2D; (b2) vertices and material frames enabling geometric quantities (e.g., curvature/dihedral) via DDG operators; (b3) coupling DDG with mechanics—energy assembly and force/Hessian evaluation—to obtain the equations of motion (EOMs).
    (c–g) Applications: mechanics studies -- (c1) coiling of rods~\cite{jawed2014coiling}, (c2) propulsion of helical filaments~\cite{jawed2015propulsion}; natural systems -- (d1) leaf shape~\cite{liang2009shape}, (d2) bacterial navigation~\cite{huang2020numerical}; flexible electronics -- (e1) wireless sensors~\cite{jang2017self}, (e2) nanotubes' pattern~\cite{k2018patterns}; robotic manipulation -- (f1) knot tying~\cite{tong2024dlodeployment}, (f2) paper folding~\cite{choi2025folding}; soft robotics -- (g1) rolling robots~\cite{huang2020dynamic}, (g2) jumping robots~\cite{huang2019highly}.} 
\label{fig:overview}
\end{figure}

Yet the same attributes that enable these capabilities also make slender structures profoundly challenging to understand and predict. 
Their behavior is governed by strongly nonlinear mechanics, including bending, twisting, stretching, shearing, and snap-through instabilities, often compounded by contact, friction, and geometric constraints~\cite{huang2025tutorial}. 
These effects are highly coupled and sensitive to boundary conditions, geometry, and material properties, producing rich behaviors such as coiling \cite{jawed2014coiling,bergou2010discrete}, knotting \cite{johanns2024capsizing, aymon2025self, tong2023snap}, and spontaneous morphogenesis \cite{moulton2013morphoelastic, lessinnes2017morphoelastic, moulton2020morphoelastic}, as shown in Fig.~\ref{fig:overview}.
Accurately capturing these phenomena is vital not only for uncovering fundamental physical principles but also for designing advanced materials~\cite{wang2025towards, bordiga2024automated}, biomedical devices~\cite{la2022flexible}, and robotic systems~\cite{choi2024dismech}.

Despite decades of progress, simulating the mechanics of flexible structures remains a challenge.
Finite element method (FEM) software (e.g., COMSOL, Abaqus, and etc.) employs shape functions and variational formulations to approximate continuum fields with high accuracy, making them the standard tool for structural analysis~\cite{multiphysics1998introduction, abaqus2011abaqus}. 
However, FEM often encounters difficulties under extreme deformations, where mesh distortion, numerical instabilities, and loss of geometric fidelity compromise robustness~\cite{cremonesi2020state}. 
Simplified alternatives, including mass–spring methods~\cite{terzopoulos1987elastically, baraff2023large, provot1995deformation} and position-based dynamics~\cite{muller2007position, macklin2016xpbd}, offer computational efficiency but typically at the expense of physical realism and predictive capability~\cite{yin2021modeling}.
As the demand for simulating flexible structures grows across disciplines, from robotics to biophysics, there is a pressing need for geometric and structure-preserving approaches that can robustly capture nonlinear mechanics across scales in an efficient manner.

To address these challenges, discrete differential geometry (DDG) has emerged as a promising framework for simulating flexible structures~\cite{grinspun2006discrete} (see Fig. \ref{fig:overview}). 
Originally developed in the computer graphics and geometry processing communities, DDG is now gaining recognition in structural mechanics for its ability to preserve geometric properties while efficiently modeling thin elastic structures such as rods, ribbons, plates, and shells~\cite{crane2018discrete}. 
Unlike conventional numerical methods that discretize governing differential equations, DDG begins by discretizing the geometry itself.
This approach allows system energies to be expressed directly in terms of the positions of mesh vertices, rather than through high-order shape functions used to approximate deformation fields.
As a result, DDG captures geometric nonlinearities in a natural way while mitigating the numerical instabilities that often affect traditional methods such as FEM~\cite{wardetzky2007discrete}.
Compared to simplified alternatives such as mass-spring systems and position-based dynamics, DDG maintains a principled connection to continuum mechanics, ensuring both physical fidelity and predictive capabilities~\cite{bergou2008discrete, grinspun2003discrete}.
By combining the rigor of continuum formulations with the robustness and efficiency needed for large-deformation and real-time simulations, DDG offers a uniquely powerful framework for modelling flexible slender structures.

In computer graphics, DDG-based simulations (e.g., Discrete Elastic Rods~\cite{bergou2008discrete}, Discrete Elastic Shell~\cite{grinspun2003discrete}, Discrete Viscous threads~\cite{bergou2010discrete}, and etc) have already achieved notable success in producing realistic animations of cloth~\cite{house2000cloth, choi2005research, bridson2005simulation}, hair~\cite{bertails2006super, kaufman2014adaptive, derouet2013inverse}, and soft machines~\cite{choi2024dismech, li2025harness}.
Building on these successes, researchers have utilized DDG-based simulations to explore a series of mechanics problems, including coiling of rods~\cite{jawed2014coiling, k2018patterns, jawed2015geometric}, flagella bundling~\cite{huang2021numerical, jawed2015propulsion, jawed2017dynamics}, knot tangling~\cite{jawed2015untangling, tong2023snap}, and elastic instabilities~\cite{huang2020shear, huang2020numerical, tong2021automated}, which also, in turn, further validate the physical accuracy of those simulations.
In bioinspired design, DDG-based models have been used to study the navigation of bacteria (e.g., buckling of flagella)~\cite{jawed2015propulsion}. 
In device applications, DDG has provided predictive simulations of flexible electronics and stretchable materials~\cite{ding2024unravelling, jung2025entanglement, huang2024integration}. %
In robotics, DDG-based simulations have provided accurate models of soft structures, which in turn facilitate the development of precise manipulation strategies for deformable objects~\cite{tong2024dlodeployment, choi2025folding} and control strategies for soft robots~\cite{choi2024dismech, huang2020numerical, huang2020dynamic}.
These diverse successes demonstrate how DDG has evolved from its origins in graphics into a versatile framework with transformative potential for engineering applications.

Given the growing successes of DDG-based simulation, its ecosystem remains far less mature than established computational tools such as FEM. 
Many of the existing works are distributed across the mechanics, robotics, and graphics communities, limiting accessibility and broader adoption. To bridge this gap, this survey aims to introduce DDG-based simulation to a broader mechanics and engineering audience, consolidate existing progress to encourage the growth of a unified ecosystem, and outline its benefits, challenges, and future opportunities.

In this review, Sec.~\ref{sec:fundamentals} outlines fundamental concepts of DDG; Sec.~\ref{sec:ModelEstablishment} reviews established DDG-based simulation models; Sec.~\ref{sec:applications} surveys applications across various engineering domains; Sec.~\ref{sec:Future_Directions} explores future opportunities; and Sec.~\ref{sec:conclusion} concludes with key perspectives.

\section{Fundamentals of Discrete Differential Geometry}
\label{sec:fundamentals}

Discrete differential geometry (DDG) provides a framework for representing smooth geometric objects, such as curves, surfaces, and manifolds, through discrete counterparts while preserving their fundamental geometric invariants~\cite{grinspun2006discrete}. 
In contrast to conventional numerical methods that approximate deformation fields with shape functions, DDG begins by discretizing the geometry itself. 
This geometry-first perspective allows quantities such as curvature, twist, and differential operators to be defined directly on meshes, ensuring that essential geometric structures remain intact even at the discrete level~\cite{huang2025tutorial}.

For one-dimensional objects, DDG represents geometry through discrete nodes connected by edges that approximate a smooth centerline. 
Edge lengths and turning angles define local bending, while a discrete Bishop material frame describes orientation along the rod~\cite{bergou2008discrete, bergou2010discrete}.
This formulation underlies the Discrete Elastic Rods (DER) model, a cornerstone of DDG in mechanics that enables accurate and efficient simulation of filaments, cables, and tendrils undergoing large deformations~\cite{jawed2018primer}.

For two-dimensional manifolds, DDG discretizes smooth surfaces into triangular meshes while retaining geometric properties such as mean and Gaussian curvature~\cite{crane2018discrete, meyer2003discrete}. 
Discrete curvature can be computed via Voronoi area weights~\cite{meyer2003discrete} or Laplace–Beltrami operators~\cite{ wardetzky2007discrete,desbrun1999implicit, pinkall1993computing}, which provide consistent approximations of smooth counterparts.
Building on these operators, the discrete shell model expresses stretching and bending energies directly in terms of vertex positions and edge lengths~\cite{grinspun2003discrete, bergou2007tracks}. 
Such formulations capture non-linear behaviors including wrinkling, buckling, and morphing, and they provide the foundation for DDG-based simulations of thin elastic plates and shells.

A unifying mathematical framework within DDG is Discrete Exterior Calculus (DEC), which generalizes vector calculus to discrete settings~\cite{desbrun2005discrete, hirani2003discrete}. 
DEC represents physical and geometric quantities as discrete differential forms on simplicial meshes: scalars on vertices (0-forms), circulations on edges (1-forms), fluxes on faces (2-forms), and volumes on cells (3-forms)~\cite{desbrun2006discrete}.
Operators such as gradient, curl, divergence, and Laplacian are realized through incidence matrices and the discrete Hodge star, ensuring that algebraic identities are preserved exactly~\cite{crane2018discrete}.
In mechanics, this structure-preserving framework supports robust geometric operators used in mechanics and facilitates stable computation on highly deformed~\cite{boom2022geometric}.
Since the geometry-preserving operators of DDG naturally support energy-based formulations of mechanics, strain energies (e.g., bending, stretching, etc.) can be expressed directly on mesh primitives (vertices, edges, faces) without the need for high-order interpolation functions~\cite{bergou2008discrete, grinspun2003discrete}. 
For slender or nearly incompressible systems, this geometry-based representation can accurately capture large-deformation behaviors and can mitigate locking effects that challenge standard FEM discretizations~\cite{saloustros2021accurate,arnold2006finite}.

Together, these principles, including geometry first discretization, Frenet-like frames for rods, curvature operators for surfaces, DEC for differential operators, and energy-based formulations, form the foundation of DDG. 
They provide a mathematically rigorous and physically faithful framework for modelling flexible structures and serve as the building blocks for the representative DDG-based simulations discussed in Section~\ref{sec:ModelEstablishment}.
A Tutorial on how to apply the DDG approach to various engineering problems has recently been published, together with open-source MATLAB codes \cite{huang2025tutorial}.

\section{DDG-Based Models for Flexible Structures}
\label{sec:ModelEstablishment}

We survey DDG-based models for flexible structures in 1D (rods/ribbons) and 2D (plates/shells). The review is organized by a geometry-first pipeline: discrete kinematics, discrete operators, energy assembly, solvers, and contact/friction. The readers can refer to~\cite{huang2025tutorial} for detailed implementation of representative DDG simulations.

\subsection{1D rod and ribbon model}

\begin{figure}
    \centering
	\includegraphics[width=0.95\textwidth]{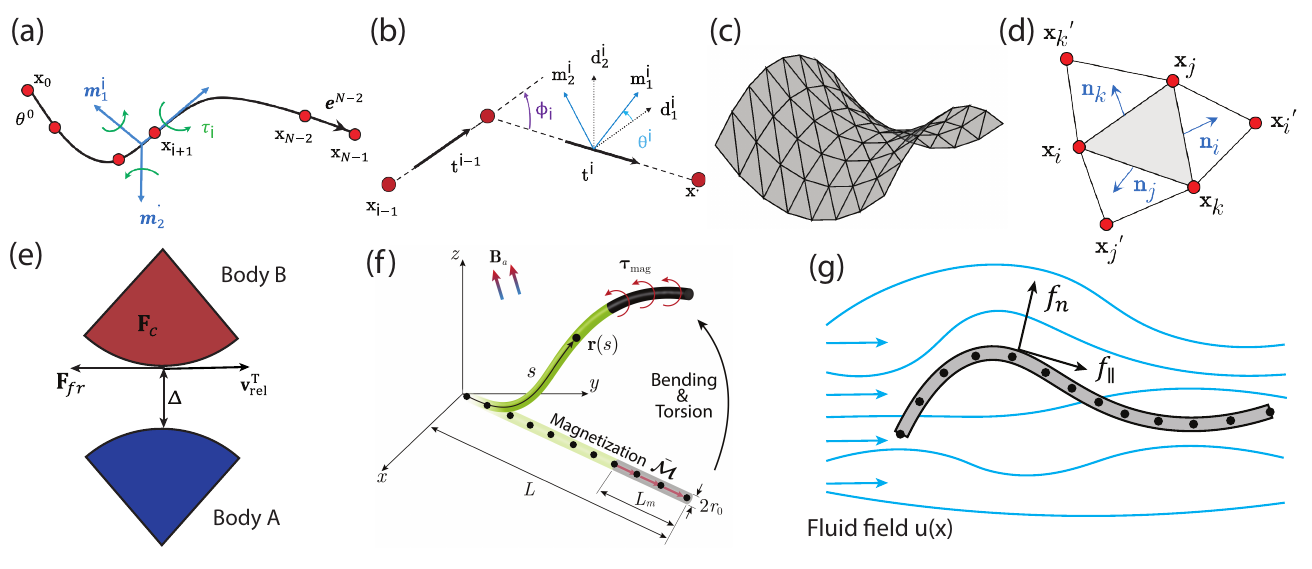}
	\caption{DDG notations and operators for 1D rods, 2D surfaces, and multi-physics modelling including frictional contact, magneto-elastic, and fluid-solid interaction. (a) Geometry of a discrete 1D structure: the centerline is described by nodes $\mathbf x_i$; material directors $\mathbf m_1^i$, $\mathbf m_2^i$ and the tangent $\mathbf t^i$ encode bending curvature $\kappa_{1, i}$, $\kappa_{2, i}$ and twist $\tau_i$.
    (b) Discrete curvature operator: curvature is obtained from the turning angle between consecutive tangents $\mathbf t^{i-1}$ and $\mathbf t^i$ (parallel transport formulation);
    (c) Discrete schematic of a 2D structure;
    (d) Mid-edge operator for 2D meshes: curvature on a triangulated surface is evaluated using mid-edge quantities defined on neighboring faces.
    (e) Contact schematic for a pair of bodies with signed distance gap $\Delta$ and tangential relative velocity $\mathbf v_\textrm{rel}$. Frictional contact responses $\mathbf F_c$ and $\mathbf F_{fr}$ can be computed based on those.
    (f) Illustration of the deformation of a magnetized filament.
    (g) The schematic of fluid-solid interaction for a filament in the fluid fields.
    } 
\label{fig:numerical}
\end{figure}

In this subsection, we introduce the DDG-based numerical formulation for 1D rod-like objects \cite{ bergou2010discrete, jawed2018primer,bergou2007tracks}.
As shown in Fig.~\ref{fig:numerical}(a), the rod centerline is discretized into $ N $ nodes: $ \{ \mathbf{x_0}, ..., \mathbf{x}_{i} ..., \mathbf{x}_{N-1} \} $
The edge vector is $\mathbf{e}^{i} = \mathbf{x}_{i+1} - \mathbf{x}_{i}$, and its length is $|| \mathbf{e}^{i} ||$. 
The Voronoi length associated with each node is the average of the two consecutive edges $\Delta {l}_{i} = \left( \lVert \mathbf{e}^{i} \rVert + \lVert \mathbf{e}^{i+1} \rVert \right) / 2$. 
Each edge, $ \mathbf{e}^{i} $, has an orthonormal adapted reference frame $ \left\{\mathbf{d}^{i}_{1}, \mathbf{d}^{i}_{2}, \mathbf{t}^{i} \right\} $ and a material frame $ \left\{\mathbf{m}^{i}_{1}, \mathbf{m}^{i}_{2}, \mathbf{t}^{i}\right\} $; both the frames share the tangent $\mathbf{t}^{i} = \mathbf e^i / \lVert \mathbf e^i \rVert$ as one of the directors.
The reference frame is updated at each time step through parallel transport in time, and, referring to Fig.~\ref{fig:numerical}(b), the material frame can be obtained from a scalar twist angle $ \theta^{i} $.
Node positions and twist angles constitute the $ 4N-1 $ sized DOF vector, $\mathbf{q} = \left[\mathbf{x}_{0}, \theta^{0}, \mathbf{x}_{1}, ..., \mathbf{x}_{N-2}, \theta^{N-2}, \mathbf{x}_{N-1} \right]$, of the discrete rod.

Based on this kinematic representation, we discuss the formulation of strains, energies, and internal elastic forces. 
An elastic rod is modeled as a mass-spring system, with a lumped mass at each node, and associated discrete stretching, bending, and twisting energies.
The stretching strain is the uniaxial elongation of each edge, and the $i$-th stretching element is $ \mathcal{S}_{i}: \{ \mathbf{x}_{i}, \mathbf{x}_{i-1}\}$.
The strain can be obtained as
\begin{equation}
\epsilon^i = \frac{\| \mathbf e^i \|}{\| \bar{\mathbf{e}}^i \|} -1.
\label{eq:StrechingStain}
\end{equation}
The $i$-th bending element is $ \mathcal{B}_{i}: \{ \mathbf{x}_{i-1}, \theta^{i-1} \mathbf{x}_{i}, \theta^{i}, \mathbf{x}_{i+1} \}$, and the associated bending strain is captured by the curvature binormal, which measures the misalignment between two consecutive edges,
\begin{equation}
(\mathbf{\kappa b})_{i} = \frac {2 \mathbf{e}^{i-1} \times \mathbf{e}^{i} } { \| \mathbf{e}^{i-1} \| \| \mathbf{e}^{i} \| + \mathbf{e}^{i-1} \cdot \mathbf{e}^{i} },
\end{equation}
and the inner products between the curvature binormal and material frame vectors give the material curvatures,
\begin{equation}
\begin{aligned}
\kappa_{1,i} & = \frac{1}{2} \left( \mathbf m_2^{i-1} + \mathbf m_2^i \right) \cdot (\kappa \mathbf b)_i, \\
\kappa_{2,i} & = - \frac{1}{2} \left( \mathbf m_1^{i-1} + \mathbf m_1^i \right) \cdot (\kappa \mathbf b)_i.
\end{aligned}
\label{eq:bendingStrain}
\end{equation}
The twisting element is the same as the bending element, and the twisting strain of $i$-th element is
\begin{equation}
\tau_{i} = \theta^{i} - \theta^{i-1} + {\delta m}_{i},
\label{TwistingCurvature}
\end{equation}
where $ \delta {m}_{i} $ is the reference twist associated with the reference frame~\citep{bergou2008discrete}.
For a rod with Young's modulus $E$, shear modulus $G$, and isotropic circular cross section, the elastic energies -- stretching, bending, and twisting --  are given by~\cite{bergou2008discrete,bergou2010discrete}
\begin{equation}
\begin{aligned}
E_s^{r} &= \frac {1} {2} \sum_{i=0}^{N_{s}} EA (\epsilon^{i})^2 || \bar{\mathbf{e}}^{i} || \\
E_b^{r} &= \frac {1} {2} \sum_{i=0}^{N_{b}} \frac {EI} { \Delta {l}_{i} } \left[ (\kappa_{1,i} - \bar{\kappa}_{1,i})^2 + (\kappa_{2,i} - \bar{\kappa}_{2,i})^2 \right] \\
E_t^{r} &= \frac {1} {2} \sum_{i=0}^{N_{t}} \frac {GJ} { \Delta {l}_{i} } ( \tau_{i} )^2,
\end{aligned}
\label{eq:totalEnergy}
\end{equation}
where $A$ is the area of cross-section, $I$ is the area moment of inertia, $J$ is the polar moment of inertia.
The case of a non-circular cross-section (e.g., ribbons) can be included in the above formulation with minor changes~\cite{bergou2008discrete,bergou2010discrete}.

The implicit Euler integration is used to solve the following $4N-1$ equation of motions and update the DOF vector $\mathbf q$ and its velocity (time derivative of DOF) $\mathbf v = \dot{\mathbf q}$ from time step $ t_{k} $ to $ t_{k+1} = t_{k} + h$ ($ h $ is the time step size):
\begin{equation}
\begin{aligned}
\mathbb{M} \Delta \mathbf{q}_{k+1} &- h \mathbb{M} \mathbf{v}_{k} - h^2 \left( \mathbf{F}^{\text{int}}_{k+1} + \mathbf{F}^{\text{ext}}_{k+1} \right) = \mathbf 0
\label{eq:implicitEulerA}
\\
\mathbf{q}_{k+1} &= \mathbf{q}_{k} + \Delta \mathbf{q}_{k+1} \\
\mathbf{v}_{k+1} &= \frac {1} {h} \Delta \mathbf{q}_{k+1},
\end{aligned}
\end{equation}
where $\mathbf{F}^{\text{int}} $ is the internal elastic force and can be derived based on the gradient of the total elastic energy,
\begin{equation}
\mathbf{F}^{\text{int}} = - \frac { \partial } {\partial \mathbf{q} } \left( E_s^{r} + E_b^{r} + E_t^{r} \right),
\end{equation}
$ \mathbf{F}^{\textrm{ext}}$ is the external force vector (e.g. gravity and damping force), $ \mathbb{M} $ is the diagonal mass matrix comprised of the lumped masses, $\dot{()}$ represents derivative with respect to time, and the superscript $k+1$ (and $k$) denotes evaluation of the quantity at time $t_{k+1}$ (and $t_k$).

\subsection{2D plate and shell model}

In this subsection, we introduce the DDG-based numerical formulation for 2D plate/shell objects \cite{seung1988defects,chen2018physical}.
As shown in Fig.~\ref{fig:numerical}(c), the 2D plate/shell object's mid-surface is discretized into $ N $ nodes: $ \{ \mathbf{x_0}, ..., \mathbf{x}_{i} ..., \mathbf{x}_{N-1} \} $, with $i  \in [0, N-1]$, and $N_{t}$ triangular meshes.
The $3N$-sized degrees of freedom (DOF) vector comprises all nodal positions, i..e., $\mathbf{q} = \left[\mathbf{x}_{0}, \mathbf{x}_{1}, ..., \mathbf{x}_{N-2}, \mathbf{x}_{N-1} \right]$.
The $\alpha$-th  stretching element is based on the triangular mesh and is comprised of three nodes, $  \mathcal{S}_{\alpha}: \{ \mathbf{x}_{i}, \mathbf{x}_{j}, \mathbf{x}_{k} \}$.
The first fundamental form for this triangular element is
\begin{equation}
\mathbb{A}_{\alpha} =
\begin{bmatrix}
(\mathbf{x}_{j} - \mathbf{x}_{i}) \cdot (\mathbf{x}_{j} - \mathbf{x}_{i}) 
& (\mathbf{x}_{j} - \mathbf{x}_{i}) \cdot (\mathbf{x}_{k} - \mathbf{x}_{i}) \\[6pt]
(\mathbf{x}_{k} - \mathbf{x}_{i}) \cdot (\mathbf{x}_{j} - \mathbf{x}_{i}) 
& (\mathbf{x}_{k} - \mathbf{x}_{i}) \cdot (\mathbf{x}_{k} - \mathbf{x}_{i})
\end{bmatrix}.
\end{equation}
The $\alpha$-th bending element of comprised of $4$ triangules and $6$ nodes , $  \mathcal{B}_{\alpha}: \{ \mathbf{x}_{i}, \mathbf{x}_{j}, \mathbf{x}_{k}, \mathbf{x}_{i}^{'}, \mathbf{x}_{j}^{'}, \mathbf{x}_{k}^{'} \}$, and the  average normal is for opposite node is $\{ \mathbf{n}_{i}, \mathbf{n}_{j}, \mathbf{n}_{k}\}$, thus the second fundamental form for this triangular element is
\begin{equation}
\mathbb{B}_{\alpha} = 2
\begin{bmatrix}
(\mathbf{n}_{j} - \mathbf{n}_{i}) \cdot (\mathbf{x}_{j} - \mathbf{x}_{i}) 
& (\mathbf{n}_{j} - \mathbf{n}_{i}) \cdot (\mathbf{x}_{k} - \mathbf{x}_{i}) \\[6pt]
(\mathbf{n}_{k} - \mathbf{n}_{i}) \cdot (\mathbf{x}_{j} - \mathbf{x}_{i}) 
& (\mathbf{n}_{k} - \mathbf{n}_{i}) \cdot (\mathbf{x}_{k} - \mathbf{x}_{i})
\end{bmatrix}.
\end{equation}
The total elastic energy is 
\begin{equation}
\begin{aligned}
E_{s}^{p} &= \sum_{\alpha =0}^{N_{t}} = \frac {1} {8} h  \; W^2_{\mathrm{SV}} \left[ \bar{\mathbb{A}}^{-1}_{\alpha} \mathbb{A}_{\alpha} - \mathbb{I} \right] \;  \sqrt{\det(\bar{\mathbb{A}}_{ijk})} \\
E_{b}^{p} &= \sum_{\alpha =0}^{N_{t}} = \frac {1} {24} h^3  \; W^2_{\mathrm{SV}} \left[ \bar{\mathbb{A}}^{-1}_{\alpha} (\mathbb{B}_{\alpha} - \bar{\mathbb{B}}_{\alpha} ) \right]  \; \sqrt{\det(\bar{\mathbb{A}}_{ijk})}
\end{aligned}
\end{equation}
where $\mathbb{I} \in \mathcal{R}^{3\times 3}$ is the identity matrix, and $W_{\mathrm{SV}}$ is the Saint Venant's linear elastic energy,
\begin{equation}
W_{\mathrm{SV}} \left[ \mathbb{K} \right]  = \frac{E \nu} {2(1 - \nu^2)} \mathrm{Tr}^2 (\mathbb{K}) +  \frac{E} {2(1+\nu)} \mathrm{Tr}(\mathbb{K}^2).
\end{equation}
The internal elastic force is related to the gradient of the total elastic energies, and the discrete-time stepping scheme for a 2D system is identical to that for the previous 1D rod-like system.

\subsection{Multi-physics modelling in DDG-based simulations}\label{s:ndim equations}

In this subsection, we review DDG-based formulations for multiphysics modeling—an essential facet of flexible structures and a major source of their nonlinear and non-smooth behavior. Leveraging a unified variational framework, DDG integrates disparate physical effects on the same discrete geometry. We illustrate this through representative couplings: frictional contact, magneto-elastic actuation, and fluid–solid interaction.

\vspace{0.5em}
\noindent\textbf{Frictional contact.}  
Frictional contact is ubiquitous in slender and flexible structures and remains one of the primary sources of nonlinearity and non-smoothness. Three major families of contact handling are commonly used. Impulse-based methods apply instantaneous impulses to prevent interpenetration~\cite{spillmann2008adaptive}, but explicit force handling often induces jitter at practical time steps. Constraint-based approaches treat normal contact and Coulomb friction as non-smooth complementarity problems~\cite{jean1987dynamics,alart1991mixed}, offering high accuracy at the cost of introducing multipliers and additional unknowns.

Penalty-energy (barrier) methods have emerged as especially well-suited for DDG-based simulations~\cite{choi2021implicit,patil2020topological,tong2023fully,li2020incremental}. They encode contact as smooth, differentiable energies whose gradients yield contact forces, preserving differentiability and enabling robust implicit time integration. For a contact gap $\Delta$ and tolerance $\delta$, a typical contact energy is:
\begin{equation}
    E_c(\Delta, \delta d) = K 
    \begin{cases}
        \Delta^2, & \Delta \leq -\delta \\
        \left(\dfrac{\delta}{15}\log(1 + \exp(-15 \dfrac{\Delta}{\delta}))\right)^2, & \Delta \in (-\delta,\delta) \\
        0, & \Delta \geq \delta,
    \end{cases}
\end{equation}
with contact forces $\mathbf F^{\mathrm{con}} = -\frac{\partial E_c}{\partial \Delta}\frac{\partial \Delta}{\partial \mathbf q}$.  
The derivative $\partial \Delta/\partial \mathbf q$ depends on the chosen distance primitive (node–node, node–edge, etc.)~\cite{li2020incremental}.  
Friction is regularized similarly. A smooth transition factor $\gamma$ mitigates the non-smooth sticking–sliding transition based on the tangential relative velocity $\mathbf v_{\mathrm{rel}}^T$:
\begin{equation}
    \gamma\left(\lVert \mathbf v_{\mathrm{rel}}^T \rVert, \nu \right) = \frac{2}{1 + \exp\left(-15 \frac{\lVert \mathbf v_{\mathrm{rel}}^T \rVert}{\nu}\right)} - 1,
\end{equation}
leading to a differentiable frictional force:
\begin{equation}
    \mathbf F^{\mathrm{Fr}} = -\mu \gamma \, \hat{\mathbf v}_{\mathrm{rel}}^T ||  \mathbf F^{\mathrm{con}}||,
\end{equation}
where $\hat{\mathbf v}_{\mathrm{rel}}^T = {\mathbf v}_{\mathrm{rel}}^T / || {\mathbf v}_{\mathrm{rel}}^T ||$ is the unit vector of the tangential relative velocity.
Because all terms are differentiable with respect to the generalized coordinates $\mathbf q$, this formulation is compatible with gradient-based simulation, optimization, and control.

\vspace{0.5em}
\noindent\textbf{Magneto-elastic coupling.}  
Magnetic actuation offers an efficient, untethered means to drive soft robots and flexible structures~\cite{zhao2019mechanics, wang2020hard}. In a DDG-based setting, magneto-elastic coupling is naturally incorporated by augmenting the system’s total potential with magnetic energy terms derived from the Maxwell stress or dipole–field interaction~\cite{tong2025real, huang2023discrete, yan2022comprehensive, sano2022kirchhoff}.  
For a body with a magnetization density $\boldsymbol{\mathcal{M}}$ in an external field $\mathbf B$, the magnetic potential can be expressed as:
\begin{equation}
    E_m = - \int_\Omega \boldsymbol{\mathcal{M}} \cdot \mathbf B \, dV.
\end{equation}
In discrete form, this integral reduces to nodal or elemental contributions based on the mesh geometry. Variations of the total magnetic potential yield the nodal-based force vector $\mathbf F^{\mathrm{mag}} = -\partial E_m/\partial \mathbf q$, which are seamlessly combined with elastic forces in the global equilibrium equations.  
This variational formulation allows magnetic and elastic forces to be treated uniformly within the same solver, enabling simulation of magnetically actuated beams, shells, or continuum robots undergoing large deformations.
The details of the numerical implementation can be referred to Ref.~\cite{huang2025tutorial}.

\vspace{0.5em}
\noindent\textbf{Fluid–solid interaction.}
Fluid–solid interaction (FSI) is central to the locomotion and manipulation of flexible structures in viscous media. Yet, DDG-based simulation work on FSI remains comparatively sparse. Existing relevant studies have concentrated largely on ciliar and flagellar systems and swimmer prototypes, reflecting the dominance of low–Reynolds-number regimes and slender-body kinematics in this domain~\cite{lim2022fabrication}. 
 
At low Reynolds numbers ($\mathrm{Re} \sim 10^{-6}\!-\!10^{-2}$), the dynamics of slender structures are governed by the strong coupling between geometric nonlinearity and viscous Stokes flow~\cite{powers2010dynamics}. Because inertia is negligible, hydrodynamic forces arise primarily from linear viscous stresses and can be described by fundamental singularity solutions (Stokeslets, dipoles, and rotlets) or their regularized counterparts~\cite{kim2013microhydrodynamics}. Within a DDG-based framework, three main classes of models are commonly used, trading off between physical fidelity, computational cost, and geometric compatibility.

Resistive–force theory (RFT) is the classical and most widely used hydrodynamic model for slender filament~\cite{gray1955propulsion}. It approximates the local drag per unit length as linearly proportional to the local velocity $\mathbf u$ relative to the surrounding fluid,

\begin{equation}
\begin{aligned}
\mathbf{F}^{\mathrm{Flu}}_{n} &= - \eta C_{n} \mathbf{u}_{n} \mathrm{ds}, \\
\mathbf{F}^{\mathrm{Flu}}_{t} &= - \eta C_{t} \mathbf{u}_{t} \mathrm{ds}, 
\end{aligned}
\end{equation}
where $\eta$ is the fluid viscosity and $C_n$, $C_t$ are normal and tangential drag coefficients determined empirically, and $\mathrm{ds}$ is the local ccross-section area.
This local formulation is computationally efficient and easy to integrate with discrete structural models, making it especially attractive for soft robotic design and control.
RFT has been successfully extended to include wall effects~\cite{clarke2006three}, non-Newtonian media~\cite{riley2017empirical}, and even granular flows~\cite{maladen2009undulatory, ding2012mechanics, maladen2011mechanical}.
However, because the hydrodynamic force at a given location depends only on local velocity, RFT cannot capture long-range interactions, bundling phenomena, or hydrodynamic coupling between adjacent segments.

In summary, these presented approaches provide a hierarchy of models for DDG-based FSI simulation. They enable the simulation and optimization of slender, flexible robots interacting with viscous environments.

\section{Applications of DDG-based Simulations}
\label{sec:applications}

Building on the foundations and representative models presented in the previous sections, DDG has increasingly emerged as a practical tool for tackling engineering problems dominated by nonlinear mechanics.
Its simulation frameworks have been used to investigate the behavior of slender and thin structures, to uncover principles in bioinspired and natural systems, to guide the design of advanced materials and devices, and to enhance modeling and control in robotics.
In the following subsections, we review representative applications across these domains, emphasizing how DDG advances both fundamental understanding and engineering innovation.

\subsection{Mechanics of Flexible Structures}

Studying the mechanics of flexible structures is important for both science and engineering, but remains challenging because their behaviors, such as bending, twisting, wrinkling, buckling, and self-contact, are highly nonlinear.
These nonlinear responses govern the functionality and failure of slender systems across scales, from DNA filaments to aerospace shells.
At the 2018 Solvay workshop on the \textit{Mechanics of slender structures in physics, biology, and engineering: from failure to functionality}, it was emphasized that discrete differential geometry (DDG) provides a more intuitive and structure-preserving framework for understanding the crucial geometric nonlinearities underlying such systems~\cite{reis2018mechanics}.

Buckling is central to slender-structure mechanics: it marks the onset of instability, sets load-bearing limits, and enables functional shape change in natural and engineered systems. 
Discrete Elastic Rods (DER) framework~\cite{bergou2008discrete} has become a cornerstone for studying such behaviors in rod-like structures, offering accurate predictions of instability onset and post-buckling evolution. 
For example, rod deposition on a moving substrate exhibits a rich set of coiling patterns captured by DER and tabletop experiments~\cite{jawed2014coiling}. 
In elastic strips, symmetry considerations explain whether snap-through proceeds via saddle-node or pitchfork bifurcations~\cite{radisson2023elastic}, and small boundary asymmetries can be transiently amplified to bias the transition pathway~\cite{wang2024transient, giudici2025transient}.
Furthermore, robotic testing with DER probes stability of helical filaments in loading scenarios~\cite{tong2021automated}.

\begin{figure}[h]
    \centering
	\includegraphics[width=0.95\textwidth]{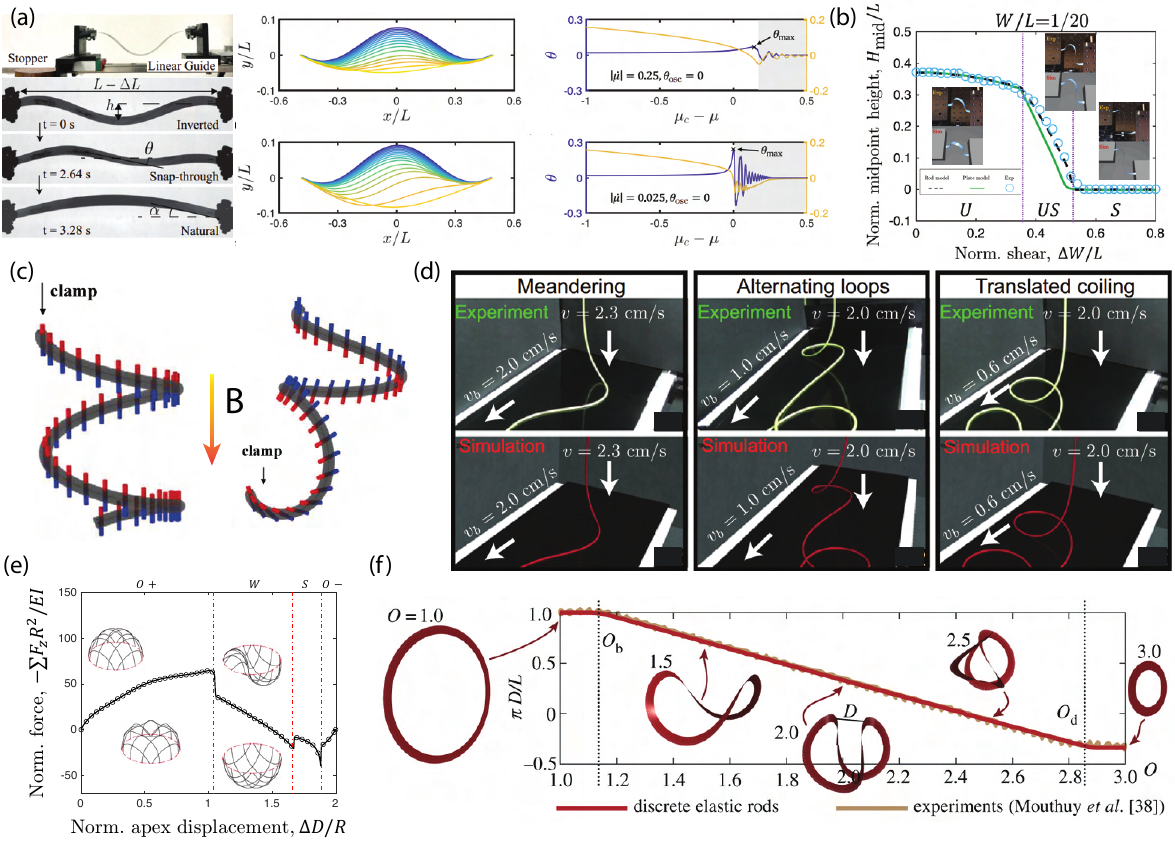}
	\caption{Applications of DDG-based simulation to mechanics study. DDG models enable quantitative analysis of nonlinear behaviors across rods, ribbons, and shells. Examples include: (a) buckling of an elastic strip and the effects of asymmetric boundary conditions on its transition~\cite{wang2024transient, giudici2025transient};  (b) bifurcation-driven shape evolution of a ribbon~\cite{huang2020shear}; (c) magneto-elastic coupling in hard-magnetic helical filaments~\cite{sano2022kirchhoff}; (d) pattern selection in rods deposited on a moving substrate~\cite{jawed2014coiling}; (g) indentation-induced buckling of a hemispherical gridshell~\cite{huang2022numerical} and (e) equilibrium shapes of an over-curved elastic ring~\cite{korner2021simple}.    
   } 
\label{fig:mechanics}
\end{figure}

The DER formulation has been extended in several directions to broaden its scope. 
First, self-contact with friction enables quantitative studies of knot mechanics, including tangling and inversions of overhand knots~\cite{tong2023snap}. 
Second, coupling to external fields—e.g., magneto-elastic effects—captures field-driven instabilities in hard-magnetic rods~\cite{sano2022kirchhoff,huang2023discrete,huang2023modeling,zhang2025achieving}. 
Furthermore, a static (equilibrium) variant supports bifurcation and stability analyses, providing a DDG-based surrogate to continuation packages such as AUTO~\cite{AUTOsite} and COCO~\cite{COCOdsweb} for tracing nonlinear equilibrium branches~\cite{huang2023bifurcations}.

Beyond rod-like structures, DDG-based simulations also play a central role in analyzing other slender systems. 
For example, the anisotropic DER was employed to study the bifurcation of thin elastic ribbons and strips, including pitchfork bifurcations and snap-through bifurcations \cite{huang2024integration, huang2024integration, huang2021snap}.
The linear Kirchhoff model was also extended to both the nonlinear Sadowsky model and the Wunderlich model \cite{korner2021simple, moore2015computation, charrondiere2024merci, huang2022discrete, neukirch2021convenient, audoly2021one}, to consider the width effect and thus bridge the gap between the 1D rod and 2D plate.
Besides 1D rods and ribbons, the nonlinear mechanics (e.g., buckling and snapping) of hollow gridshells \cite{huang2021numerical, huang2022numerical, baek2019rigidity, baek2018form, panetta2019x} and axisymmetric shells \cite{huang2024discrete, huang2024snap} were also extensively investigated through a DDG-based approach. 
The mechanics of a 2D surface were studied through a discrete shell model~\cite{seung1988defects}, including gut growth ~\cite{savin2011growth}, leaf configuration ~\cite{liang2009shape}, band bifurcation \cite{huang2020shear}, and bilayer systems \cite{chen2018physical}.

In addition to forward simulation, DDG also supports inverse design, which is an important problem in mechanics. Li et al. propose a biomimetic ``Turing machine'' for morphing bilayer ribbons~\cite{li2025biomimetic}, develop a DDG-based solver tailored to bilayers~\cite{li2025harness}, and introduce an inverse elastica to recover undeformed configurations, validated against DER~\cite{li2025inverse}.


\subsection{Natural and Bioinspired Systems}

Nature abounds with slender and thin structures, for example, plant tendrils that coil around supports~\cite{goriely1998spontaneous}, epithelial tissues that fold into villi and crypts~\cite{hannezo2011instabilities}, or leaves that wrinkle and curl as they grow~\cite{liang2009shape}.
Understanding their mechanics is vital for both revealing the physical principles behind the biological form and translating those principles into efficient, adaptive designs.
DDG provides an effective framework for modeling the nonlinear behaviors of slender structures found in nature and offers valuable insights for bioinspired systems.

\begin{figure}[h]
    \centering
    \includegraphics[width=0.75\textwidth]{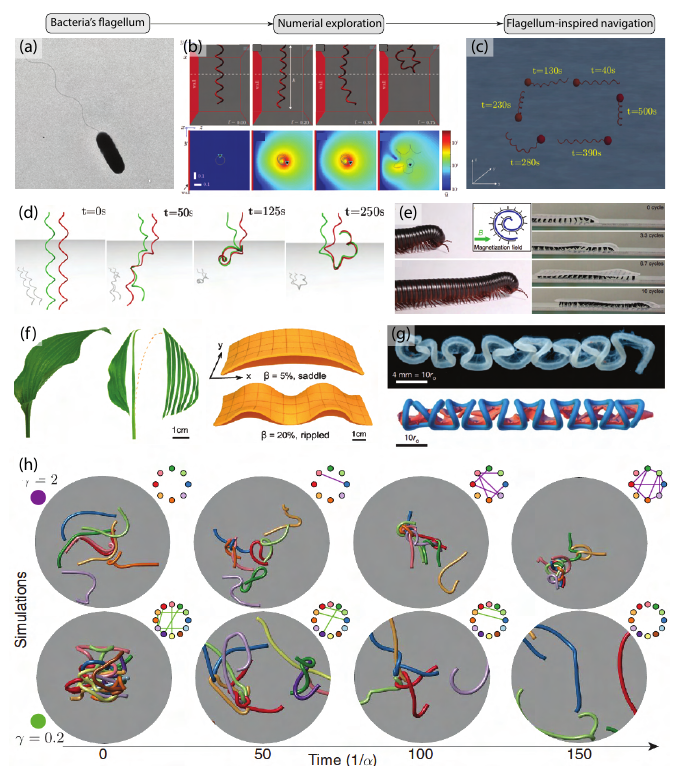}
    \caption{Applications of DDG-based simulation to natural systems. 
    DDG models reveal biomechanical mechanisms underlying bioinspired phenomena. 
    Shown are: (a) rotation of a bacterial flagellum in a viscous medium (image adapted from~\cite{yang2024ceanimonas}); 
    (b) DDG-based simulation of flagellar dynamics~\cite{jawed2017dynamics}; 
    (c) navigation and control leveraging elastic–hydrodynamic coupling~\cite{huang2020numerical}; 
    (d) bundling under simultaneous rotation of multiple flagella~\cite{tong2023fully}; 
    (e) ciliary locomotion~\cite{gu2020magnetic}; 
    (f) leaf morphogenesis from differential growth~\cite{liang2009shape}; 
    (g) gut looping driven by mesentery–tube growth mismatch~\cite{savin2011growth}; 
    (h) emergent entanglement–disentanglement in worm collectives~\cite{patil2023ultrafast}.}
    \label{fig:natural}
\end{figure}

Slender filaments represent a canonical system for investigating bioinspired mechanics.
A prime example is the locomotion of bacterial flagella~\cite{van1665observations}. 
Bacteria use their helical flagella to navigate through fluid environments, where propulsion arises from solid–fluid interaction forces~\cite{lim2022fabrication}.
This coupling gives rise to nonlinear behaviors such as buckling~\cite{son2013bacteria}, bundling~\cite{kim2003macroscopic}, and tumbling~\cite{macnab1977normal}, which in turn generate distinct motion patterns. 
A series of studies~\cite{jawed2015propulsion, jawed2017dynamics, huang2020numerical, tong2023fully} have employed DDG-based simulations to investigate these phenomena in detail, offering insights into both the mechanics of flagellar propulsion and the design of bioinspired locomotion systems. 
Beyond flagella, Patil et al.~\cite{patil2023ultrafast} applied a DDG-based framework to study the rapid entanglement and disentanglement of worm collectives, highlighting how self-organization in natural systems can inspire the design of adaptive, reconfigurable materials and robotic assemblies.

DDG-based simulations have also been used to study shell-like biological membranes.
Savin et al.~\cite{savin2011growth} combined experiments, physical mimics, and a DDG-based framework to show that gut looping morphogenesis arises from differential growth between the gut tube and its supporting mesentery, with predictions consistent across chick, quail, finch, and mouse intestines. Similarly, Liang and Mahadeva~\cite{liang2009shape} demonstrated that the saddle-like and rippled shapes of long leaves result from elastic relaxation due to differential growth, supported by both experiments and generalized plate theory with a DDG-based numerical simulation. Together, these studies illustrate how DDG-based models can capture the essential mechanics of biological morphogenesis while providing design principles for bioinspired thin structures and deployable devices.

\subsection{Functional Structures and Devices}

Beyond studying mechanics and natural systems, DDG-based simulations are increasingly valuable for understanding and designing functional structures that exploit geometric nonlinearity for performance.
Flexible mechanical metamaterials provide a unifying context in which geometry, rather than composition, governs response.
Prior work has shown how mechanism-driven layouts (origami, kirigami), instability-enabled lattices (snapping and post-buckling pathways), and topological architectures yield tunable, programmable behaviors~\cite{rafsanjani2019programming,zhang2017printing}.

\begin{figure}[!h]
    \centering
	\includegraphics[width=0.95\textwidth]{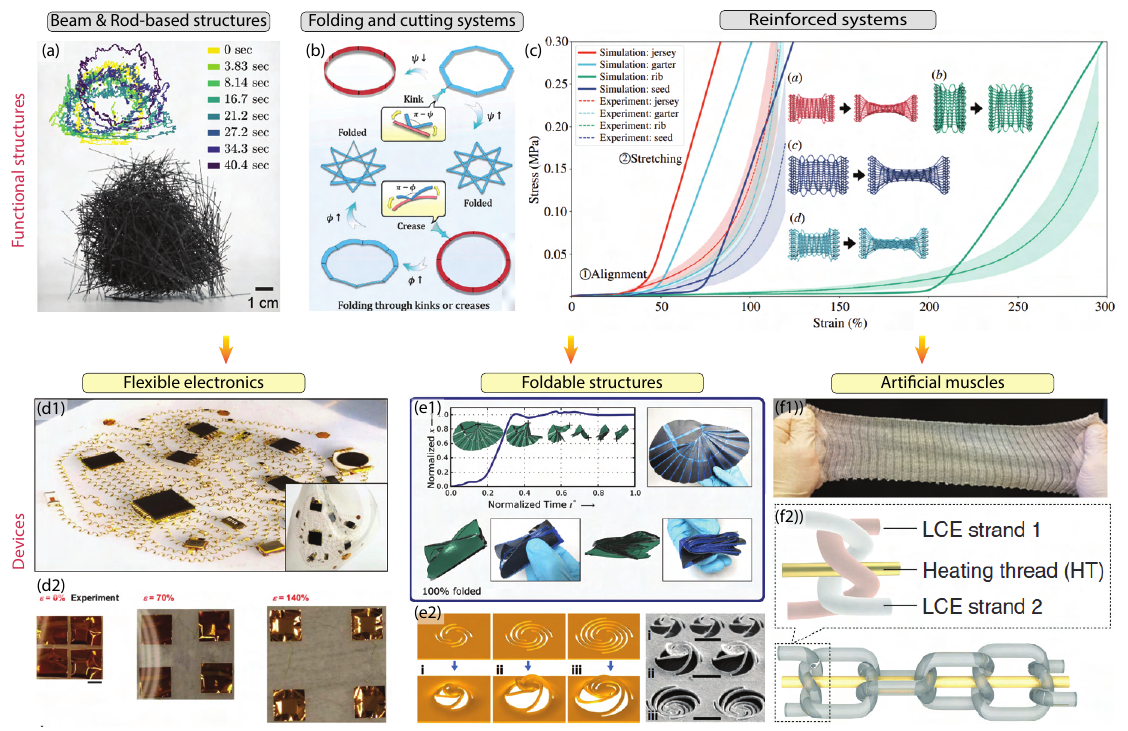}
	\caption{DDG-based simulations for functional structures and opportunities for device concepts.
    DDG-based simulations have shown accuracy on representative functional structures: (a) beam and rod systems, exemplified by rod packaging~\cite{jung2025entanglement}; (b) folding and cutting systems, exemplified by meta-ribbons~\cite{huang2024integration}; (c) topology-reinforced systems, exemplified by knitted structures~\cite{ding2024unravelling}. These structures motivate devices that were not analyzed with DDG in the cited works but are strong candidates for DDG-based modeling: flexible electronics, including (d1) stretchable electrical interconnects~\cite{jang2017self} and (d2) serpentine microstructure stretchable electronics~\cite{zhang2013buckling}; foldable systems, including (e1) kirigami-inspired foldable wings~\cite{faber2018bioinspired} and (e2) reconfigurable opto-NEMS kirigami~\cite{chen2021electromechanically}; artificial muscles, including (f1) Lyocell-based textile actuators~\cite{maziz2017knitting} and (f2) knotted liquid crystal elastomers(LCE) actuators~\cite{chen2024knotted}.
   } 
\label{fig:devices}
\end{figure}

We first review studies that already employ DDG-based simulation on functional structures closely related to device building blocks.
As shown in Fig.~\ref{fig:devices}, three representative works capture core architectural behaviors.
X-ray tomography combined with DDG-based rod simulations reveals an entanglement transition in random rod packings and maps a stability–rigidity phase space relevant to textiles and architected bundles~\cite{jung2025entanglement}.
A hierarchical yarn-level DDG model explains the nonlinear elasticity, anisotropy, and programmability of knitted fabrics~\cite{ding2024unravelling}.
A DDG framework for meta-ribbons shows how kinks and creases coordinate folding pathways through distinct bifurcation mechanisms, enabling programmable meta-structures~\cite{huang2024integration}.

Moving from structures to devices, most recent analyses still rely on finite element methods.
In mechanical metamaterials, finite element models have enabled multistep transformations, reprogrammable behaviors, and information encoding~\cite{meng2020multi,meng2023encoding,meng2022deployable}.
In flexible electronics, finite element analysis underpins ultrastretchable serpentine interconnects~\cite{zhang2013buckling}, deterministic three-dimensional assembly via controlled buckling in skin-mounted sensors~\cite{jang2017self}, and curvature-programmed bilayer ribbon networks for inverse-designed mesostructures that integrate sensing and optoelectronic functions~\cite{shen2024curvature}.
In foldable structures, finite element models support the design and characterization of kirigami- and origami-inspired deployables, from bioinspired foldable wings~\cite{faber2018bioinspired} to reconfigurable optoelectromechanical kirigami~\cite{chen2021electromechanically}.
In artificial muscles, finite element analysis has been used to model textile actuators~\cite{maziz2017knitting}, while, notably, Chen et~al.~\cite{chen2024knotted} leverages DDG-based simulation to optimize knot types for designing knotted liquid-crystal–elastomer actuators.

DDG offers a strong alternative for device-level modeling.
By discretizing geometry directly, it preserves discrete curvature and strain operators, yields consistent energies, forces, and tangents, and handles large deformations, bifurcations, and contact via mesh-consistent gap functions.
These properties align with the dominant mechanisms in devices such as stretchable electrical interconnects, kirigami-inspired deployables, and textile- or LCE–based actuators.
Although the cited device exemplars were primarily analyzed with finite element methods, they are natural targets for predictive DDG-based simulation.
In regimes dominated by geometric nonlinearity, self-contact, and pattern selection—where structure preservation and robust tangents are critical—DDG can credibly replace FEM or serve as a drop-in alternative, with the added benefit of clean differentiability for large-deformation modeling, inverse design, and control.
This progression, from validated DDG results in functional structures to device opportunities, is reflected in Fig.~\ref{fig:devices}.

\subsection{Robotics}

\subsubsection{Manipulation of Deformable Structures}

Though DDG-based simulations have gained significant traction within the graphics and mechanics communities, their adoption in robotics has historically been less than anticipated. 
This gap may be attributed to the limited number of robotics researchers familiar with DDG-based simulation tools, despite recent efforts to integrate them into more canonical rigid-body robotics simulators such as MuJoCo~\cite{chen2025mujoco_der}. 
Nevertheless, in recent years the use of DDG-based simulation in robotics has grown rapidly with one of the key applications being the usage of DER for solving the robotic manipulation of deformable linear objects (DLOs).

\begin{figure}[h]
    \centering
	\includegraphics[width=0.95\textwidth]{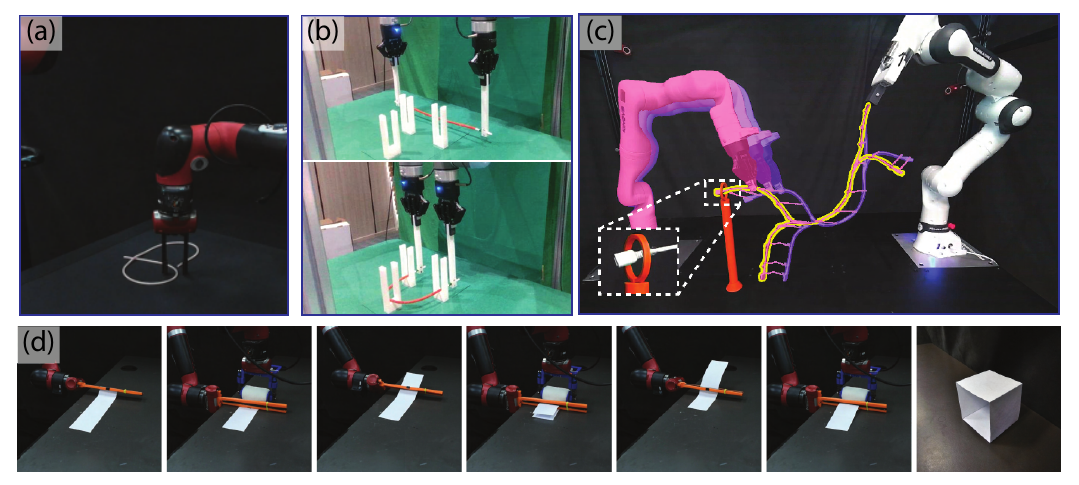}
	\caption{DDG-based simulation applications in robotic manipulation of deformable objects. (a) Single arm DLO pattern deployment to achieve open-loop knot tying~\cite{tong2024dlodeployment}. (b) Bimanual cable manipulation in constrained 3D environments~\cite{yu2025wholebody_dlo_manipulation}. (c) Branched DLO shape forming in end insertion~\cite{chen2025deft}.
    (d) Square paper origami achieved by precise repeated symmetrical paper folding~\cite{choi2025folding} . 
    } 
\label{fig:ddg_robot_fig}
\end{figure}

Among the many manipulation problems, DLO deployment—the task of laying a DLO onto a surface in a prescribed pattern—is particularly challenging due to non-linear interaction of contact and elasticity. 
Early work by Lv et al.~\cite{lv2019shape_forming} introduced an empirical position-based pattern matching controller to deploy simple shapes (e.g., semicircles, sine waves) using a single manipulator, though their results were limited to simulation. 
This was later extended to dual-arm manipulation in follow-up work~\cite{lv2021flexiblecables}, though this work was still solely within simulation. 
A subsequent unification of these approaches~\cite{lv2022dlo_manipulation} validated both single and dual-arm deployment strategies on real robots and even incorporated frictional contact to the DER simulation rather than relying on the assumption of perfect sticking. 
Building on this line of research, Tong et al.~\cite{tong2024dlodeployment} proposed a data-driven sim-to-real neural controller capable of deploying DLOs into arbitrary patterns with a single arm. 
Their controller generalized across material properties by leveraging the Buckingham $\pi$ theorem, and real robot experiments demonstrated open-loop deployment with such high accuracy that the system could repeatedly tie knots directly from deployed patterns (Fig.~\ref{fig:ddg_robot_fig}b).

Beyond deployment, researchers have also investigated dexterous manipulation of DLOs in constrained 3D environments (Fig.~\ref{fig:ddg_robot_fig}a). 
Yu et al.~\cite{yu2025wholebody_dlo_manipulation} used DER as a global planner to generate coarse feasible solutions for bimanual manipulation tasks, which were then executed with a Jacobian-based local controller. 
Similarly, Choi et al.~\cite{choi2025folding} employed DER as a reduced-order model to simulate symmetric paper folding (Fig.~\ref{fig:ddg_robot_fig}c). 
They trained neural force manifolds from data generated from this model and computed optimal folding trajectories by traversing the learned manifolds while minimizing predicted forces.
Similar to~\cite{tong2024dlodeployment}, this work also leveraged Buckingham $\pi$ theorem to achieve generalization across material properties.

Another emerging direction has been the use of DER in perception tasks. Early work by Javdani et al.~\cite{javdani2011dlo_perception} employed a DER-inspired energy model to estimate physical parameters of DLOs from real data. 
More recently, Choi et al.~\cite{choi2023mbest} proposed a method for 2D instance segmentation of binary DLO masks, resolving crossings by minimizing a DER-inspired cumulative discrete bending energy objective. 
This method was later integrated into~\cite{tong2024dlodeployment} to rapidly generate discretized deployment patterns. 
Sun et al.~\cite{sun2024dlo_3d_perception} tackled DLO 3D reconstruction by segmenting 2D images, combining them with point cloud data, and then smoothing the reconstruction using DER, taking advantage of its implicit bending energy minimization.
The high repeatability of robot manipulators has also enabled researchers to use robots as tools for mechanics studies. 
For instance, Tong et al.~\cite{tong2021automated} used a robot manipulator to test the stability of stiff DLOs with helical centerlines, comparing experimental results against DER simulations.

Finally, significant attention has been devoted to improving the speed and accuracy of DER simulations. 
Chen et al.~\cite{chen2024differentiable} introduced DDER, a differentiable variant of DER, and used it to develop DEFORM, a hybrid framework that couples DDER with a neural network to compensate for integration errors, achieving accurate mid-air bimanual DLO shape forming. 
Furthermore, efficient parameter identification was achieved thanks to framework's differentiability. 
Building on this, Chen et al.~\cite{chen2025deft} extended their approach to branched DLOs (BDLOs) for automating wire harness assembly, combining differentiable simulation with learning-based corrections tailored for branched structures (Fig.~\ref{fig:ddg_robot_fig}d).

\subsubsection{Soft Robotics Design and Control}






Building on the use of DDG-based simulations for manipulating deformable objects with rigid manipulators, a second major robotics application arises in soft robotics, where the simulator models the robot’s deformable body itself. Soft robotics has emerged as a powerful paradigm for designing adaptive, safe, and versatile machines. Their inherent compliance enables safe physical interaction with humans, operation in unstructured or confined environments, and manipulation of delicate or irregular objects~\cite{rus2015design, kim2013soft}. These capabilities underpin a wide range of applications, including biomedical devices~\cite{gu2023soft,kim2019ferromagnetic}, search-and-rescue systems~\cite{hawkes2017soft}, wearable technologies~\cite{ding2018human}, and autonomous exploration~\cite{chen2025sppiro,fan2024vacuum,fan2025sparc}.

\begin{figure}[h]
    \centering
    \includegraphics[width=0.95\textwidth]{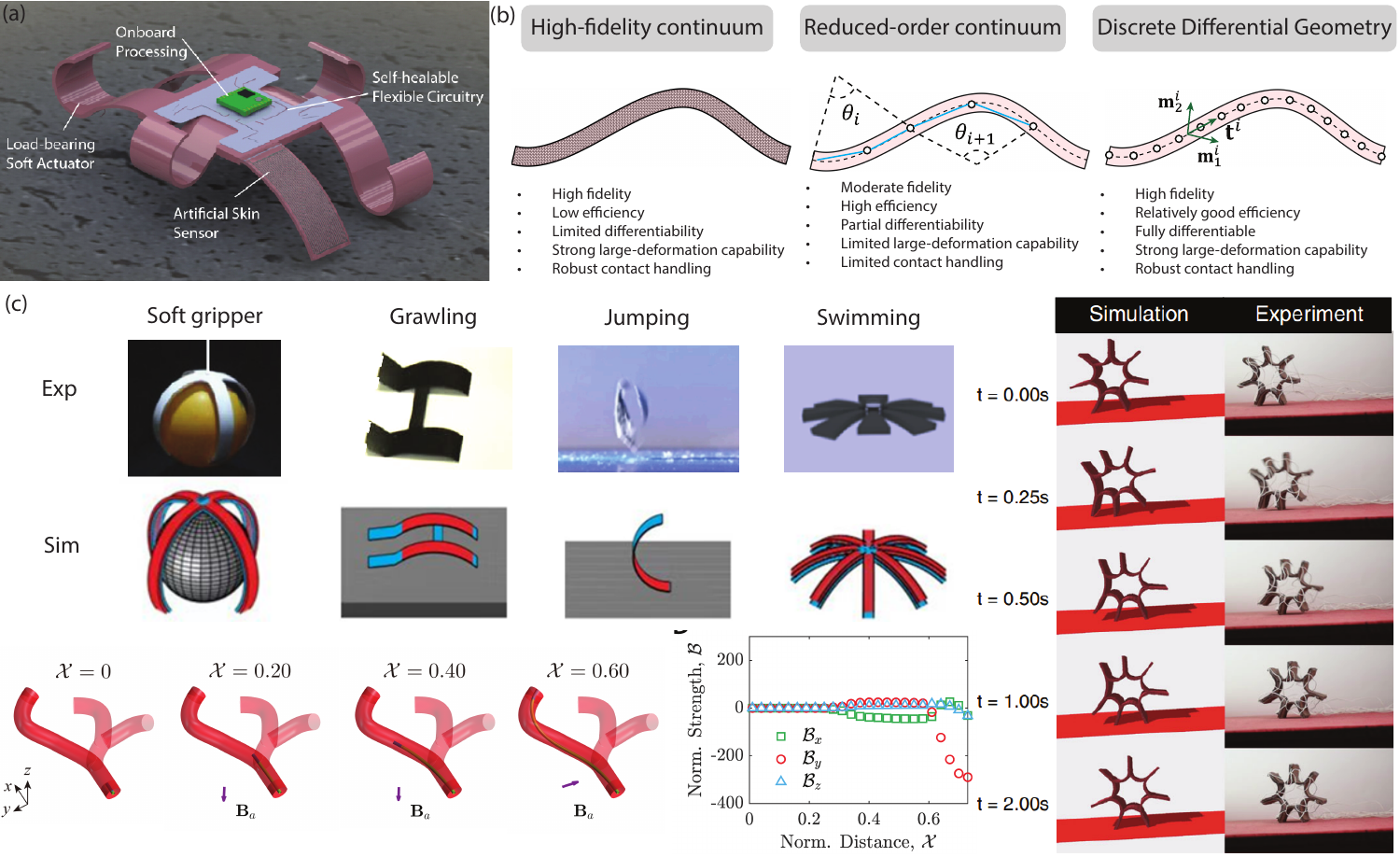}
    \caption{DDG-based simulation for robotic manipulation of deformable objects.
    (a) Soft robots that leverage compliant bodies for actuation, sensing, and control to navigate complex environments.
    (b) Three modeling paradigms: high-fidelity continuum (e.g., FEM), reduced-order continuum (e.g., piecewise-constant curvature, PCC), and discrete differential geometry (DDG); key pros/cons summarized.
    (c) DDG-based simulation across soft-robot archetypes: soft grippers, terrestrial/legged locomotion, wall-crawling/pipe-crawling robots, and swimming robots~\cite{li2025harness}
    (d) Closed-loop control with DDG—illustrated on a magnetic soft continuum robot (MSCR) for image-guided/clinical tasks~\cite{tong2025real}
    (e) Comparison of DDG simulation and experiment for dynamic soft-robot behaviors~\cite{huang2020dynamic}}
    \label{fig:soft_robot}
\end{figure}

Modeling such systems, however, is fundamentally challenging. Continuous deformation, nonlinear material responses, distributed actuation, frequent contact and self-contact, and strong environmental coupling yield high-dimensional, strongly nonlinear dynamics~\cite{trivedi2008soft,laschi2016soft,armanini2023soft}. Traditional approaches each face significant trade-offs. Finite element methods (FEM) provide high-fidelity solutions but are computationally expensive and difficult to differentiate~\cite{armanini2023soft}, while reduced-order models (e.g., piecewise-constant curvature approximations) achieve real-time performance but oversimplify large-deformation behaviors~\cite{webster2010design}, limiting their applicability in optimization, control, and sim-to-real transfer~\cite{hu2019chainqueen,du2021diffpd}.

DDG offers a powerful alternative. By discretizing curvature, strain, and other intrinsic quantities directly on mesh primitives, DDG preserves physical invariants and avoids common numerical artifacts~\cite{huang2025tutorial}. In practice, DDG frameworks can naturally handle large deformations, bifurcations, contact, and multi-physics interactions while remaining computationally efficient and fully differentiable—capabilities that are essential for optimization, inverse design, and closed-loop control~\cite{tong2025real,huang2020dynamic,du2022granular,hao2023tumbling,hao2024tumbling_attitude,lim2023low_reynolds,tong2025inverse}. The comparison of those three mentioned modelling methods are illustrated in Fig.~\ref{fig:soft_robot}(b).

One of the most prominent areas where these strengths have been demonstrated is bioinspired locomotion. DDG-based simulation has been used to model a star-shaped soft robot rolling over curved terrain and to study gait selection across varying substrate geometries~\cite{huang2020dynamic}. As illustrated in Fig.~\ref{fig:soft_robot}(c), DDG frameworks have also been applied to simulate multiple locomotion modes—crawling~\cite{dong2019multi}, jumping~\cite{xu2022insect}, and swimming~\cite{yin2021visible}—within a unified bilayer modeling environment~\cite{li2025harness}. These approaches have directly informed robot design: for example, DDG-based models guided the fabrication and performance analysis of a star-shaped swimmer~\cite{huang2021star_swimming} and a frog-inspired soft swimmer~\cite{huang2022swimming2d}, enabling systematic studies of navigation strategies in fluid environments. Similarly, a bacteria-inspired soft robot operating near an air–fluid interface was designed and validated using DDG-based simulations~\cite{du2021airfluid}. For flagellated systems, DDG-based studies have captured low-Reynolds-number locomotion and control phenomena, including bi-flagellated swimmers capable of tumbling-based attitude adjustments~\cite{hao2023tumbling} and regime transitions arising from viscous–inertial interplay~\cite{lim2023low_reynolds}.

Beyond locomotion, DDG-based simulations are also increasingly integrated into design and control pipelines. Their differentiability and computational efficiency make them particularly suitable for model-predictive control (MPC) and learning-based approaches. For instance, Tong et al.~\cite{tong2025real} combined DDG-based simulation with numerical differentiation to evaluate how control inputs, such as external magnetic fields, influence the configuration of a magnetic continuum soft robot (MSCR). This approach enabled real-time MPC for navigating confined lumens without wall contact during minimally invasive procedures. Furthermore, DDG-based simulations can serve as efficient data generators for training data-driven controllers. Tong et al.~\cite{tong2025inverse} developed a learning-based pipeline that leverages DDG-generated data to explore parameter spaces for a bioinspired jumping robot~\cite{wang2023insect}, identifying optimal design parameters for targeted performance. Together, these studies highlight the growing role of DDG not only in modeling but also in the design and control of soft robots.

While many existing DDG-based simulators were initially designed for specific embodiments or tasks, recent efforts have focused on building more general-purpose simulation platforms. The Elastica simulator~\cite{naughton2021elastica}, based on Cosserat rod theory, enables simulation of a wide range of soft structures and has been applied to model systems such as octopus-inspired arms~\cite{wang2021optimal} and dual-ring soft robots~\cite{wang2021computationally}. More recently, DisMech~\cite{choi2024dismech} has emerged as one of the first DDG-based platforms capable of modeling diverse soft robotic systems, including multi-legged walkers, active entanglement grippers, and continuum limbs with real-to-simulation validation. Building on this foundation, Mat-DisMech~\cite{lahoti2025matdismech} extends these capabilities by supporting multi-material actuation and improved contact handling, further broadening the scope of DDG-based modeling in soft robotics.

In summary, DDG-based simulations uniquely combine geometric fidelity, computational efficiency, and differentiability, overcoming the limitations of traditional methods. Their demonstrated success across fluidic, terrestrial, and hybrid soft robots—together with their suitability for inverse design, learning-based control, and real-time deployment—positions DDG as a foundational tool for the next generation of soft robotic systems.

\section{Future Directions}
\label{sec:Future_Directions}
\begin{figure}[h]
    \centering
    \includegraphics[width=0.95\textwidth]{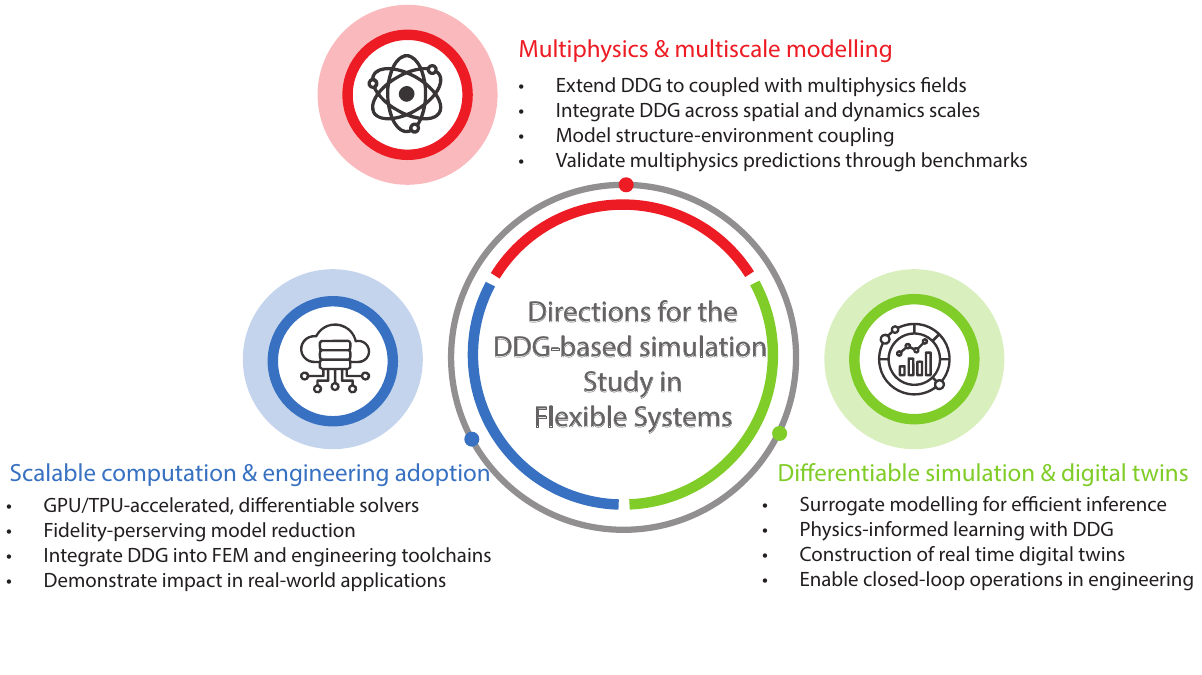}
    \caption{Future developments and research directions for DDG-based simulations in flexible systems.}
    \label{fig:future}
\end{figure}

We have reviewed the relevant work on DDG-based simulations and their applications in different engineering domains.
Two strengths stand out: a geometry-preserving formulation that promotes numerical stability and physical fidelity, and inherent differentiability that exposes gradients essential for optimization, control, and design. These advantages position DDG as a compelling foundation for next-generation modeling of flexible systems. Fig.~\ref{fig:future} summarizes three priority directions; we expand on them below.

\begin{enumerate}[label=\arabic*.]
  \item \textbf{Multiphysics and multiscale extensions.} 
  Future work must extend DDG beyond purely elastic formulations to integrate magnetic, thermal, electrostatic, and fluidic effects. Geometry-first discretizations are particularly well-suited for coupling across scales, from nanoscale biomolecules to macroscopic deployable devices, and from quasi-static morphogenesis to dynamic instabilities. Capturing structure–environment interactions, such as viscous fluid–filament dynamics or thermally driven instabilities, will broaden DDG’s reach across disciplines.

  \item \textbf{Differentiable simulation, data integration, and digital twins.}
  Because DDG models are inherently differentiable, they align naturally with optimization, inverse design, and machine learning. Embedding DDG within physics-informed neural networks, surrogate modeling frameworks, and digital twin architectures will accelerate the design and control of complex systems. Real-time updating digital twins, informed by experimental data streams, can close the loop between simulation and practice in robotics, biomedicine, and structural monitoring.

  \item \textbf{Scalable computation and engineering adoption.}
  Achieving real-time capability remains a central challenge. GPU/TPU-accelerated implementations, reduced-order yet curvature-preserving formulations, and parallelized optimization pipelines will enable interactive modeling and large-scale design exploration. Adoption by engineering practice requires open, robust software platforms, integration with established FEM pipelines, and demonstration in applications ranging from medical devices to multifunctional soft robots.
\end{enumerate}

Together, these directions define a comprehensive roadmap in which geometry-preserving modeling, data-driven intelligence, and real-world applications converge to establish DDG-based simulation as a unifying framework for the next generation of flexible system design, control, and discovery.

\section{Conclusions}
\label{sec:conclusion}

Discrete Differential Geometry has matured into a versatile framework for simulating nonlinear flexible systems. Its geometry-first perspective yields models that balance accuracy, efficiency, and robustness while remaining fully differentiable. By consolidating progress across rods, shells, and multiphysics couplings, and by demonstrating impact in mechanics, biology, functional devices, and robotics, this review positions DDG as a unifying approach for next-generation simulations. Looking ahead, advances in multiphysics extensions, differentiable digital twins, and scalable GPU implementations will accelerate its adoption in engineering design, biomedical modeling, and intelligent robotics. DDG is thus poised to become a cornerstone of flexible system simulation at the intersection of geometry, computation, and application.

\subsection*{Acknowledgments}

W.H. acknowledges the start-up funding from Newcastle University, UK.
M.L. acknowledges the start-up funding from The University of Birmingham, UK.
K.J.H. acknowledges the financial support from the Ministry of Education, Singapore, under its Academic Research Fund Tier 3 (Grant MOEMOET32022-0002).

\addcontentsline{toc}{section}{References}  
\bibliographystyle{elsarticle-num}
\bibliography{ref}

\begin{thebibliography}{100}
\bibitem{van1665observations}
A.~Van~Leewenhoeck, Observations, communicated to the publisher by mr. antony van leewenhoeck, in a dutch letter of the 9th of octob. 1676. here english'd: concerning little animals by him observed in rain-well-sea. and snow water; as also in water wherein pepper had lain infused, Philosophical Transactions (1665-1678) 12  821--831.
\end{thebibliography}

\end{document}